\documentstyle[12pt]{article}
\begin{document}
\begin{center}
{\large\bf
 Exact Renormalization Group for O(4) Gauged Supergravity\\}
\vspace{1cm}
L.N. Granda
\footnote{e-mail: Granda@quantum.univalle.edu.co}\\
{\it Departamento de Fisica, Universidad del Valle}\\
{\it A.A. 25360, Cali, Colombia}\\
and\\
S.D. Odintsov
\footnote{e-mail: Odintsov@quantum.univalle.edu.co}\\
{\it Department of Physics and Mathematics\\
Tomsk Pedagogical University\\
634041 Tomsk, Russia and}\\
{\it Departamento de Fisica, Universidad del Valle}\\
{\it A.A. 25360, Cali, Colombia}\\
\end{center}

We study exact renormalization group (RG) in O(4) gauged supergravity
using the effective average action formalism. The nonperturbative
RG equations for cosmological and newtonian coupling constants
are found. It is shown the existence of (nonstable) fixed point
of these equations. The solution of RG equation for newtonian
coupling constant is qualitatively the same as in Einstein gravity
(i.e. it is growing at large distances).

{\large\bf 1.\\}
There was recently much activity in the study of nonperturbative RG dynamics in field theory models (for  recent refs., see \cite{rg}). One of the versions of nonperturbative RG based on the effective average action has been developed in ref. \cite{r} for
 Einstein gravity (the gauge dependence problem in this formalism has been studied in ref \cite{fo}). The nonperturbative RG equations for cosmological and newtonian coupling constants have been obtained in ref. \cite{r}. The comparison between quantum correction
 to newtonian coupling from nonperturbative RG \cite{r} and from effective field theory technique \cite{d} has been done.\\
It is quite interesting to generalize nonperturbative RG (or evolution equation) to supersymmetric versions of gravity theories. The theories of such type with cosmological term are gauged supergravities (SG), which may show the way to
solve  the cosmological
 constant problem \cite{cd}. Moreover, if local supersymmetry indeed exists in nature, it most probably should be realized in its gauged form. Hence, the purpose of this work will be to study the nonperturbative evolution equation in O(4) gauged SG 
which is considered in components [1,2].\\
We follow the formalism of ref. \cite{r} developed for gravitational theories. The basic elements of it are the background field method (see \cite{bos} for a review) and the truncated nonperturbative evolution equation for the effective average action 
$\Gamma_k[g,\bar{g}]$:
\par$$
\partial_t\Gamma_k[g,\bar{g}]=\frac{1}{2}Tr\left[\left(\Gamma_k^{(2)}[g,\bar{g}]+R_k^{grav}[\bar{g}]\right)^{-1}\partial_tR_k^{grav}[\bar{g}]\right]-
$$
$$
 Tr\left[\left(-M[g,\bar{g}]+R_{k}^{gh}[\bar{g}]\right)\partial_t R_{k}^{gh}[\bar{g}]\right]
\eqno{(1)}$$
where $t=\ln k$, $k$ is the nonzero momentum scale, $R_k$ are cut-offs, $M[g,\bar{g}]$ are ghost operators, $\bar{g}_{\mu\nu}$ is the background metric, $g_{\mu\nu}=\bar{g}_{\mu\nu}+h_{\mu\nu}$ where $h_{\mu\nu}$ is the quantum field. $\Gamma_k^{(2)}$ is 
the hessian of $\Gamma_k[g,\bar{g}]$ with respect to $g_{\mu\nu}$ at fixed $\bar{g}_{\mu\nu}$ (for more details see \cite{r}).\\
Our purpose will be to get the explicit projected form of Eq. (1) to the space of low derivatives functionals in O(4) SG, i.e. to obtain the nonperturbative RG equations for cosmological and newtonian constants. Note that we work on nonsupersymmetric De 
Sitter background. The local SUSY is broken due to gauge fixing in graviton and gravitino sectors and due to the introduction of cut-off terms.\\
{\large\bf 2.\\}
We will start from the theory with the following lagrangian density \cite{das,bwt} (see also \cite{bz})
\par$$
L=-2\kappa^2R+4\kappa^2\frac{\partial_{\mu}\Phi\partial^{\mu}\Phi^*}{(1-|\Phi|^2)^2}-
32g^2\kappa^4\left(1+\frac{2}{1-|\Phi|^2}\right)+\frac{1}{8}(F_{\mu\nu}^{ij})^2+
$$
$$
\left[\frac{1}{8}\frac{\Phi}{1-\Phi^2}\left(\Phi\delta_{ik}\delta_{jl}-
\frac{1}{2}\epsilon_{ijkl}\right)\left(F_{\mu\nu}^{ij}F_{\mu\nu}^{kl}+
iF_{\mu\nu}^{ij}\tilde{F_{\mu\nu}^{kl}}\right)+h.c.\right]
$$
$$
+\frac{1}{2}\epsilon^{\mu\nu\rho\sigma}\bar{\Psi}_{\mu}^i\gamma_5\gamma_{\nu}D_{\rho}\Psi_{\sigma}^i+\frac{2\sqrt{2}\kappa g}{\sqrt{1-|\Phi|^2}}\bar{\Psi}_{\mu}^i\sigma_{\mu\nu}\Psi_{\nu}^i
$$
$$
+\frac{1}{2}\bar{\chi}^i\hat{D}\chi^i+\frac{2g\kappa}{\sqrt{1-|\Phi|^2}}
\bar{\Psi}_{\mu}^i\gamma_{\mu}\left(\Phi_1+i\gamma_5\Phi_2\right)\chi^i+...
\eqno{(2)} $$

where $\kappa^2=1/32\pi G$, quartic fermionic terms are not written explicitly in (2), $\hat{D}=\gamma_{\mu}D_{\mu}$, $D_{\mu}=\tilde{D}_{\mu}+gA_{\mu}$,
$\tilde{F}_{\mu\nu}=\frac{1}{2}\epsilon_{\mu\nu\rho\sigma}F_{\rho\sigma}$.\\
The fields content of such O(4) gauged SG includes: graviton, an O(4) gauge field, a complex scalar $\Phi=\Phi_1+i\Phi_2$, four Majorana gravitino and four Majorana spinors (for more details see \cite{das,bwt}). The action (2) is invariant under $N=4$ 
supersymmetry \cite{das,bwt,bz}.\\
We work on De Sitter background ($R_{\mu\nu}=\frac{1}{4}R g_{\mu\nu}$) and put all rest background fields to be zero. 

The corresponding one-loop calculation of the effective action has been actually done in ref. \cite{ft} (for a correspondent study on hyperbolic background, see \cite{boz}). One has only to collect the corresponding pieces to $Z=e^{-\Gamma}$, taking into 
account the Jacobians. As a result we obtain:
\par$$
Z_{\Phi}=\left[\det \triangle_0(-6g^2\kappa^2)\right]^{-1}, 
Z_A=\left[\frac{\{\det\triangle_0(0)\}^2}{\det\triangle_V(R/4)}\right]^3
$$
$$
Z_{grav}=\frac{\{\det\triangle_V(-R/4)\}}{\det^{1/2}\triangle_T(\frac{2}{3}R-V_0/2\kappa^2)\det^{1/2}\triangle_0(- V_0/2\kappa^2)} 
$$
$$
Z_{gravitino}=\frac{\det\triangle_{3/2}(m^2)}{\{\det\triangle_{1/2}(4m^2)\}}\;\;\;\mbox{,}
Z_{\chi}=\det\triangle_{1/2}(0) 
\eqno{(3)}$$
Here the expressions written in $\{...\}$ are the ghost contributions in the corresponding sector, $V_0=-36g^2\kappa^4$,
$m^2=8g^2\kappa^2$.\\
In the calculation of the graviton contribution to one-loop effective action we used Fock-de Donder gauge. For the calculation of the vector contribution the standard minimal gauge has been used. The calculation of gravitino contribution has been done in the 
gauge \cite{ft}:
\par$$
L_g=\frac{1}{4}\bar{\Psi}\gamma_0(\hat{D}+2m)^{-1}\Psi
\eqno{(4)}$$
where the gauge parameter $\gamma_0$ is taken to be zero at the end of the calculations.\\
The following  operators have been introduced:
\par$$
\triangle_0(X)\phi=(-D^2+X)\phi \;\;\mbox{,}\;\;\
\triangle_{1/2}(X)\psi=(-D^2+R/4+X)\psi 
$$
$$
\triangle_{3/2}(X)\phi_{\nu}^{\bot}=\left(-D^2_{\mu\nu}+\frac{1}{3}R g_{\mu\nu}+X g_{\mu\nu}\right)\phi_{\nu}^{\bot}
\eqno{(5)}$$
where indexes show the spin, $D^{\mu}\phi_{\mu}^{\bot}=0$. More details can be found in refs. \cite{ft,gp,cd}.
Note that the operators $\triangle_V$ and $\triangle_T$ which correspond to vector and tensor are not constrained (for details see \cite{fo})
\par$$
\det\triangle_V(X)=\det(-D^2+X)_V\;\;\mbox{,}\;\
\det\triangle_T(X)=\det(-D^2+X)_T
\eqno{(6)}$$
As one sees only gravitino contribution in (3) contains differentially constrained operator. In order to translate it to unconstrained operator one can use the following relation \cite{ft}:
\par$$
\det\triangle_{3/2}(X)=\lbrack\frac{\det(\hat{D}+\sqrt{X})_{\psi_{\mu}}}{\det(-D^2+X)_{\psi}}\rbrack^2
\eqno{(7)}$$
where
\par$$
\lbrack\det(\hat{D}+\sqrt{X})_{\psi_{\mu}}\rbrack^2=
$$
$$
\det\{-g_{\mu\nu}D^2+\frac{R}{4}
g_{\mu\nu}-\frac{R}{24}\left(\gamma_{\mu}\gamma_{\nu}-
\gamma_{\nu}\gamma_{\mu}\right)+Xg_{\mu\nu}\}
\eqno{(8)}$$
Then, the gravitino contribution to the one-loop partition function may be rewritten as
\par$$
Z_{gravitino}=\frac{\det_{3/2}\lbrack -g_{\mu\nu}D^2+\frac{R}{4} g_{\mu\nu}-
\frac{R}{24}(\gamma_{\mu}\gamma_{\nu}-\gamma_{\nu}\gamma_{\mu})+m^2g_{\mu\nu}\rbrack}{\{\det\triangle_{1/2}(4m^2)\}\lbrack\det_{1/2}(-D^2-\frac{R}{4}+m^2)\rbrack^{-2}}
\eqno{(9)}$$
Collecting all above pieces one can write the effective average action (we suppose that cut-offs in all sectors are chosen to be the same)
\par$$
\bar{\Gamma}_k[g,g]=\frac{1}{2}Tr_T\ln\left[
Z_{Nk}(-D^2+\frac{2}{3}R-\frac{1}{2\kappa^2} V_0+k^2R^{(0)})\right]
$$
$$
+\frac{1}{2}Tr_0\ln\left[ Z_{Nk}\left(-D^2-\frac{1}{2\kappa^2} V_0+ k^2 R^{(0)}\right)\right]
$$
$$
-Tr_V\ln\left[ -D^2-\frac{R}{4}+k^2R^{(0)}\right]
+Tr_0\ln\left[ Z_{Nk}\left(-D^2+\frac{1}{6\kappa^2}V_0+k^2R^{(0)}\right)\right]
$$
$$
+3Tr_V\ln\left[ Z_{Nk}\left(-D^2+\frac{R}{4}+k^2R^{(0)}\right)\right]
-6Tr_0\ln\left[-D^2+k^2R^{(0)}\right]
$$
$$
-Tr_{1/2}\ln\left[ Z_{Nk}\left(-D^2+\frac{R}{4}+k^2R^{(0)}\right)\right]
+2Tr_{1/2}\ln\left[ Z_{Nk}\left(-D^2+m^2+k^2R^{(0)}\right)\right]
$$
$$
-Tr_{3/2}\ln\left[ Z_{Nk}\left(-g_{\mu\nu}D^2+\frac{R}{4} g_{\mu\nu}-
\frac{R}{24}\left(\gamma_{\mu}\gamma_{\nu}-\gamma_{\nu}\gamma_{\mu}
\right)+m^2g_{\mu\nu}+k^2R^{(0)}g_{\mu\nu}\right)\right]
$$
$$
+Tr_{1/2}\ln\left[-D^2+\frac{R}{4}+4m^2+k^2R^{(0)}\right]
\eqno{(10)}$$

Here $Z_{Nk}$ is renormalization function (compare with ref.[2]).\\

{\large\bf 3.\\}
Let us write the exact RG equation for the theory with the action (2). We differentiate the average effective action with respect to $t$, then we set $\bar{g}_{\mu\nu}=g_{\mu\nu}$. This corresponds to special choice of a field theory. It follows that the gauge fixing term $S_{gf}$ 
vanishes so that the LHS of the renormalization group equation reads (see \cite{r} for more details)
\par$$
\partial_t\Gamma_k[g,g]=2\kappa^2\int  d^4x\sqrt{g}\left[-R(g)\partial_t(Z_{Nk})+2\partial_t(Z_{Nk}\bar{\lambda}_k)\right]
\eqno{(11)}$$
Now we want to find the RHS of the evolution equation. To this end, we differentiate the average action (10) with respect to $t$. Then we expand the operators in (10) with respect to the curvature $R$ because we are only interested in terms of order $\int
 d^4x \sqrt{g}$ and $\int d^4x \sqrt{g}R$:
\par$$
\triangle_i^{-1}\left(aR-2b\bar{\lambda}_k+k^2R^{(0)}\right)=\triangle_i^{-1}\left(-2b\bar{\lambda}_k+k^2R^{(0)}\right)-
$$
$$
\triangle_i^{-2}\left(-2b\bar{\lambda}_k+k^2R^{(0)}\right)aR+0(R^2)
\eqno{(12)}$$
where $a$ and $b$ are the constants and $\bar{\lambda}_k=\frac{1}{4\kappa^2}V_0$. For a more compact notation let us introduce
\par$$
\tilde{\triangle}_{ib}(z)=\triangle_i\left(-2b\bar{\lambda}_k+k^2R^{0}(z)\right)
\eqno{(13)}$$
\par$$
N(z)=\frac{\partial_t\left[Z_{Nk}k^2R^{(0)}(z)\right]}{Z_{Nk}}\;\;\;\,N_0(z)=\partial_t\left[k^2R^{(0)}(z)\right]
\eqno{(14)}$$
Here the variable z replaces $-D^2/k^2$. These steps lead then to
\par$$
\partial_t\bar{\Gamma}_k[g,g]=\frac{1}{2}Tr_T\left[N\tilde{\triangle}_{T1}^{-1}\right]+\frac{1}{2}Tr_S\left[N\tilde{\triangle}_{S1}^{-1}\right]-Tr_V\left[N_0\tilde{\triangle}_{V0}^{-1}\right]
$$
$$
+Tr_S\left[N\tilde{\triangle}_{S,-1/3}^{-1}\right]+3Tr_V\left[N\tilde{\triangle}_{V0}^{-1}\right]-6Tr_S\left[N_0\tilde{\triangle}_{S0}^{-1}\right]
$$
$$
-Tr_{1/2}\left[N\tilde{\triangle}_{1/2,0}\right]+2Tr_{1/2}\left[N\tilde{\triangle}_{1/2,1/6}^{-1}\right]-Tr_{3/2}\left[N\tilde{\triangle}_{3/2,1/6}^{-1}\right]
$$
$$
+Tr_{1/2}\left[N_0\tilde{\triangle}_{1/2,2/3}^{-1}\right]-\frac{R}{4}\left\{\frac{4}{3}Tr_T\left[N\tilde{\triangle}_{T1}^{-2}\right]+Tr_V\left[N_0\tilde{\triangle}_{V0}^{-2}\right]\right.
$$
$$
\left.+3Tr_V\left[N\tilde{\triangle}_{V0}^{-2}\right]-Tr_{1/2}\left[N\tilde{\triangle}_{1/2,0}^{-2}\right]\right.
$$
$$
\left.-Tr_{3/2}\left[\left\{g_{\mu\nu}-\frac{1}{3}\sigma_{\mu\nu}\right\}N\tilde{\triangle}_{3/2,1/6}^{-2}\right]+Tr_{1/2}\left[N_0\tilde{\triangle}_{1/2,2/3}^{-2}\right]\right\}
\eqno{(15)}$$
The terms with $N_0$ are the contributions of the ghosts.\\
As a next step we evaluate the traces. We use the heat kernel expansion which for an arbitrary function of the covariant Laplacian $W(D^2)$ reads
\par$$
Tr_j\left[W(-D^2)\right]=(4\pi)^{-2}tr_j(I)\left\{Q_2[W]\int d^4x\sqrt{g}\right.
$$
$$
\;\;\left. +\frac{1}{6}Q_1[W]\int d^4x\sqrt{g}R+0(R^2)\right\}
\eqno{(16)}$$
where by $I$ we denote the unit matrix in the space of fields on which $D^2$ acts. Therefore $tr_j(I)$ simply counts the number of independent degrees of freedom of the field
The sort $j$ of fields enters (16) via $tr_j(I)$ only. Therefore we will drop the index $j$ of $\bar{\triangle}_{ja}$ after the evaluation of the traces in the heat kernel expansion.\\
The functionals $Q_n$ are the Mellin transforms of $W$,
\par$$
Q_n[W]=\frac{1}{\Gamma(n)}\int_0^{\infty}dz z^{n-1}W(z)\;\;\, (n>0)
\eqno{(17)}$$
Now we have to perform the heat kernel expansion (16) in Eq. (15). This
leads to a polynomial in $R$ which is the RHS of the evolution equation (11). By the comparison of coefficients with the LHS of the evolution equation (11) we obtain in order of $\int d^4x\sqrt{g}$
\par$$
\partial_t(Z_{Nk}\bar{\lambda}_k)=\frac{1}{4\kappa^2}\frac{1}{(4\pi)^2}\left\{
5Q_2\left[N\tilde{\triangle}_1^{-1}\right]-10Q_2\left[N_0\tilde{\triangle}_0^{-1}\right]+8Q_2\left[N\tilde{\triangle}_0^{-1}\right]\right.
$$
$$
\left.+Q_2\left[N\tilde{\triangle}_{-1/3}^{-1}\right]-8Q_2\left[N\tilde{\triangle}_{1/6}^{-1}\right]+4Q_2\left[N_0\tilde{\triangle}_{2/3}^{-1}\right]\right\}
\eqno{(18)}$$
and in order of $\int d^4x\sqrt{g}R$
\par$$
\partial_t Z_{Nk}=-\frac{1}{24\kappa^2}\frac{1}{(4\pi)^2}\left\{10Q_1\left[N\tilde{\triangle}_1^{-1}\right]-36Q_2\left[N\tilde{\triangle}_1^{-2}\right]\right.
$$
$$
\left.-20Q_1\left[N_0\tilde{\triangle}_0^{-1}\right]-12Q_2\left[N_0\tilde{\triangle}_0^{-2}\right]+2Q_1\left[N\tilde{\triangle}_{-1/3}^{-1}\right]\right.
$$
$$
\left. -24Q_2\left[N\tilde{\triangle}_0^{-2}\right]+16Q_1\left[N\tilde{\triangle}_0^{-1}\right]-16Q_1\left[N\tilde{\triangle}_{1/6}^{-1}\right]\right.
$$
$$
\left.+48Q_2\left[N\tilde{\triangle}_{1/6}^{-2}\right]-12Q_2\left[N_0\tilde{\triangle}_{2/3}^{-2}\right]+8Q_1\left[N_0\tilde{\triangle}_{2/3}^{-1}\right]\right\}
\eqno{(19)}$$
The cutoff dependent integrals are defined in [2]
\par$$
\Phi_n^p(w)=\frac{1}{\Gamma(n)}\int_0^{\infty}dz z^{n-1}\frac{R^{(0)}(z)-zR^{(0)'}(z)}{[z+R^{(0)}(z)+w]^p}
$$
$$
\tilde{\Phi}_n^p(w)=\frac{1}{\Gamma(n)}\int_0^{\infty}dz z^{n-1}\frac{R^{(0)}(z)}{[z+R^{(0)}(z)+w]^p}
\eqno{(20)}$$
for $n>0$. It follows that $\Phi_0^p(w)=\tilde{\Phi}_0^p(w)=(1+w)^{-p}$ for
$n=0$. In addition we use the fact that
\par$$
N=\frac{\partial_t\left[Z_{Nk}k^2R^{(0)}(-D^2/k^2)\right]}{Z_{Nk}}=\left[2-\eta_N(k)\right]k^2R^{(0)}(z)+2D^2R^{(0)'}(z)
\eqno{(21)}$$
with $\eta_N(k)=-\partial_t(\ln Z_{Nk})$ being the anomalous dimension of the operator $\sqrt{g}R$. Then we can rewrite the equations (18) and (19) in terms of $\Phi$ and $\tilde{\Phi}$. This leads to the following system of equations
\par$$
\partial_t\left(Z_{Nk}\bar{\lambda}_k\right)=\frac{1}{4\kappa^2}\frac{1}{(4\pi)^2}k^4\left[10\Phi_2^1(\bar{\lambda}_k/k^2)-4\Phi_2^1(0)\right.
$$
$$
\;\;\left. +2\Phi_2^1(-\bar{\lambda}_k/3k^2)-16\Phi_2^1(\bar{\lambda}_k/6k^2)+8\Phi_2^1(\bar{2\lambda}_k/3k^2)\right.
$$
$$
\left. -\eta_N(k)\left\{5\tilde{\Phi}_2^1(\bar{\lambda}_k/k^2)+8\tilde{\Phi}_2^1(0)+\tilde{\Phi}_2^1(-\bar{\lambda}_k/3k^2)-8\tilde{\Phi}_2^1(\bar{\lambda}_k/6k^2)\right\}\right]\mbox{,}
\eqno{(22)}$$
\par$$
\partial_t(Z_{Nk})=-\frac{1}{24\kappa^2}\frac{1}{(4\pi)^2}k^2\left\{20\Phi_1^1(\bar{\lambda}_k/k^2)-72\Phi_2^2(\bar{\lambda}_k/k^2)\right.
$$
$$
\;\;\left. -8\Phi_1^1(0)-36\Phi_2^2(0)+4\Phi_1^1(-\bar{\lambda}_k/3k^2)-32\Phi_1^1(\bar{\lambda}_k/6k^2)\right.
$$
$$
\;\;\left. +96\Phi_2^2(\bar{\lambda}_k/6k^2)-24\Phi_2^2(\bar{2\lambda}_k/3k^2)+16\Phi_1^1(\bar{2\lambda}_k/3k^2)\right.
$$
$$
\left.-\eta_N(k)\left[10\tilde{\Phi}_1^1(\bar{\lambda}_k/k^2)-36\tilde{\Phi}_2^2(\bar{\lambda}_k/k^2)+2\tilde{\Phi}_1^1(-\bar{\lambda}_k/3k^2)\right.\right.
$$
$$
\left.\left.
+48\tilde{\Phi}_2^2(0)+16\tilde{\Phi}_1^1(0)-16\tilde{\Phi}_1^1(\bar{\lambda}_k/6k^2)+48\tilde{\Phi}_2^2(\bar{\lambda}_k/6k^2)\right]\right\}
\eqno{(23)}$$
Now we introduce the dimensionless, renormalized Newtonian constant and cosmological constant
\par$$
g_k=k^2G_k=k^2Z_{Nk}^{-1}\bar{G}\;\;\mbox{,}\;\;\lambda_k=k^{-2}\bar{\lambda}_k
\eqno{(24)}$$
Here $G_k$ is the renormalized Newtonian constant at scale $k$. The evolution equation for $g_k$ reads then
\par$$
\partial_t g_k=\left[2+\eta_N(k)\right]g_k
\eqno{(25)}$$
>From (23) we find the anomalous dimension $\eta_N(k)$
\par$$
\eta_N(k)=g_kB_1(\lambda_k)+\eta_N(k)g_kB_2(\lambda_k)
\eqno{(26)}$$
where
\par$$
B_1(\lambda_k)=\frac{1}{3\pi}\left[5\Phi_1^1(\lambda_k)-18\Phi_2^2(\lambda_k)-2\Phi_1^1(0)-9\Phi_2^2(0)\right.
$$
$$
\left. +\Phi_1^1(-\lambda_k/3)-8\Phi_1^1(\lambda_k/6)+24\Phi_2^2(\lambda_k/6)-6\Phi_2^2(2\lambda_k/3)+4\Phi_1^1(2\lambda_k/3)\right]
$$
$$
B_2(\lambda_k)=-\frac{1}{6\pi}\left[5\tilde{\Phi}_1^1(\lambda_k)-18\tilde{\Phi}_2^2(\lambda_k)+\tilde{\Phi}_1^1(-\lambda_k/3)\right.
$$
$$
\left. -6\tilde{\Phi}_2^2(0)+8\tilde{\Phi}_1^1(0)-8\tilde{\Phi}_1^1(\lambda_k/6)+24\tilde{\Phi}_2^2(\lambda_k/6)\right]
\eqno{(27)}$$
Solving (26)
\par$$
\eta_N(k)=\frac{g_kB_1(\lambda_k)}{1-g_kB_2(\lambda_k)}
\eqno{(28)}$$
we see that the anomalous dimension $\eta_N$ is a nonpertubative quantity. From (22) we obtain the evolution equation for the cosmological constant
\par$$
\partial_t(\lambda_k)=-\left[2-\eta_N(k)\right]\lambda_k-\frac{1}{8\pi}g_k\left[10\Phi_2^1(\lambda_k)-4\Phi_2^1(0)\right.
$$
$$
\left. +2\Phi_2^1(-\lambda_k/3)-16\Phi_2^1(\lambda_k/6)+8\Phi_2^1(2\lambda_k/3)-\right.
$$
$$
\left.-\eta_N(k)\left\{5\tilde{\Phi}_2^1(\lambda_k)+8\tilde{\Phi}_2^1(0)+\tilde{\Phi}_2^1(-\lambda_k/3)-8\tilde{\Phi}_2^1(\lambda_k/6)\right\}\right]
\eqno{(29)}$$
The equations (25) and (29) together with (28) give the system of differential equations for the two $k$-depending coupling constants $\lambda_k$ and $g_k$. These equations determine the value of the running Newtonian constant and cosmological constant at the scale $k<<
\Lambda_{cut-off}$. Above evolution equations include non-perturbative effects which go beyond a simple one-loop calculation.\\
Next, we estimate the qualitative behaviour of the running Newtonian constant as above system of RG equations is too complicated and cannot be solved analytically. To this end we assume that the cosmological constant is much smaller than the IR cut-off 
scale, $\lambda_k<<k^2$, so we can put $\lambda_k=0$ that simplify Eqs. (25) and (27). After that, we make an expansion in powers of $(\bar{G}k^2)^{-1}$ keeping only the first term (i.e. we evaluate the functions $\Phi_n^p(0)$ and $\tilde{\Phi}_n^p(0)$) and
 finally obtain (with $g_k\sim k^2\bar{G}$)
\par$$
G_k=G_o\left[1-w\bar{G}k^2+...\right]
\eqno{(30)}$$
where
\par$$
w=\frac{3}{2\pi}\Phi_2^2(0)=\frac{3}{2\pi}
\eqno{}$$
In case of Einstein gravity, similar solution has been obtained in refs. [2,3]. In getting (30) we use the same cut-off function as in [2].\\
We see that $w>0$, what means that the newtonian coupling decreases as $k^2$ increases; i.e. we find that gravitational coupling is antiscreening 
like in Einstein gravity.\\
Let us now investigate the problem of existance of critical points in the theory under investigation. We search the points at which  r.h.s. of Eqs. (25) and (29) are equal to zero. The numerical analysis of correspondent RG system 
 gives:
\par$$
\lambda_k=-0.375\;\;\;\;\mbox{,}\;\;\;\; g_k=1.36
\eqno{(31)}$$
These points actually correspond to UV unstable fixed points. Note that solutions (31) do not give the solution of cosmological constant problem as the result of non-perturbative RG running.\\
There are also fixed points with negative newtonian coupling constant.We do not consider
 these points as non-physical ones.

In summary, we discussed exact RG equations for O(4) gauged SG and
found corresponding RG equations for cosmological and newtonian couplings. 
As the approximate solution of these equations the antiscreening behaviour 
of newtonian coupling constant is obtained. The critical point of SG model
 under discussion is also evaluated.

Let us mention briefly few possible extensions of our results. First, it would 
be very interesting to study the gauge dependence problem in the models of SG. 
To do this one can use the gauge-fixing independent effective action formalism
 (like in ref.[3] for Einstein gravity) or better truncation of evolution equation
 which leads to few more nonperturbative RG equations for gauge parameters.
Second, it is desireable to generalize the effective average action formalism
using superspace background field method. Then one may hope to get explicitly
supersymmetric formulation of exact RG even at finite cut-off.

{\bf Acknowledgments}
We would like to thank S.Falkenberg for independent check of part of results presented in this work.  L.N.G. was supported by COLCIENCIAS 
(Colombia) Project No. 1106-05-393-95. S.D.O was supported in part by COLCIENCIAS.

\end{document}